\RequirePackage{ifpdf}
\ifpdf % We are running pdfTeX in pdf mode
\documentclass[pdftex]{sigma}
\else
\documentclass{sigma}
\fi

\begin{document}

\allowdisplaybreaks

\renewcommand{\thefootnote}{$\star$}

\renewcommand{\PaperNumber}{078}

\FirstPageHeading

\ShortArticleName{A Particular Solution of a Painlev\'{e} System}

\ArticleName{A Particular Solution of a Painlev\'{e} System\\ in Terms of the Hypergeometric Function $\boldsymbol{{}_{n+1}F_n}$\footnote{This paper is a
contribution to the Special Issue ``Relationship of Orthogonal Polynomials and Special Functions with Quantum Groups and Integrable Systems''. The
full collection is available at
\href{http://www.emis.de/journals/SIGMA/OPSF.html}{http://www.emis.de/journals/SIGMA/OPSF.html}}}

\Author{Takao SUZUKI}

\AuthorNameForHeading{T.~Suzuki}

\Address{Department of Mathematics, Kobe University, Rokko, Kobe 657-8501, Japan}
\Email{\href{mailto:suzukit@math.kobe-u.ac.jp}{suzukit@math.kobe-u.ac.jp}}

\ArticleDates{Received June 23, 2010, in f\/inal form September 29, 2010;  Published online October 07, 2010}

\Abstract{In a recent work, we proposed the coupled Painlev\'{e} VI system with $A^{(1)}_{2n+1}$-symmetry, which is a higher order generalization of the sixth Painlev\'{e} equation ($P_{\rm{VI}}$).
In this article, we present its particular solution expressed in terms of the hypergeometric function ${}_{n+1}F_n$.
We also discuss a degeneration structure of the Painlev\'{e} system derived from the conf\/luence of ${}_{n+1}F_n$.}


\Classification{17B80; 33C20; 34M55}

\section{Introduction}

The main object in this article is the coupled Painlev\'{e} VI system with $A^{(1)}_{2n+1}$-symmetry given in \cite{FS,Su}, or equivalently, the Hamiltonian system $\mathcal{H}_{n+1,1}$ given in \cite{T}.
It is expressed as a~Hamiltonian system on $\mathbb{P}^1(\mathbb{C})$
\begin{gather}\label{Eq:CP6}
	t(t-1)\frac{dq_i}{dt} = \frac{\partial H}{\partial p_i},\qquad
	t(t-1)\frac{dp_i}{dt} = -\frac{\partial H}{\partial q_i},\qquad i=1,\ldots,n,
\end{gather}
with
\begin{gather*}
	H  = \sum_{i=1}^{n}H_{\rm{VI}}\left[\sum_{j=0}^{n}\alpha_{2j+1}-\alpha_{2i-1}-\eta,\sum_{j=0}^{i-1}\alpha_{2j},
\sum_{j=i}^{n}\alpha_{2j},\alpha_{2i-1}\eta;q_i,p_i\right]\\
\phantom{H=}{} + \sum_{1\leq i<j\leq n}(q_i-1)(q_j-t)\{(q_ip_i+\alpha_{2i-1})p_j+p_i(q_jp_j+\alpha_{2j-1})\},
\end{gather*}
where $H_{\rm{VI}}$ is the Hamiltonian for $P_{\rm{VI}}$ def\/ined as
\begin{gather*}
	H_{\rm{VI}}[\kappa_0,\kappa_1,\kappa_t,\kappa;q,p]  = q(q-1)(q-t)p^2 - \kappa_0(q-1)(q-t)p\\
\phantom{H_{\rm{VI}}[\kappa_0,\kappa_1,\kappa_t,\kappa;q,p]  =}{}
- \kappa_1q(q-t)p - (\kappa_t-1)q(q-1)p + \kappa q.
\end{gather*}
Here $\alpha_0,\ldots,\alpha_{2n+1}$ and $\eta$ are f\/ixed parameters satisfying a relation $\sum\limits_{i=0}^{2n+1}\alpha_i=1$.
We assume that the indices of $\alpha_i$ are congruent modulo $2n+2$.
Note that the system~\eqref{Eq:CP6} includes $P_{\rm{VI}}$ as the case $n=1$.
The aim of this article is to present a particular solution of the system~\eqref{Eq:CP6} expressed in terms of the hypergeometric function ${}_{n+1}F_n$.

The hypergeometric function ${}_{n+1}F_n$ is def\/ined by the power series
\[
	{}_{n+1}F_n\left[\begin{array}{c}a_0,\ldots,a_n\\b_1,\ldots,b_n\end{array};t\right] = \sum_{i=0}^{\infty}\frac{(a_0)_i(a_1)_i\cdots(a_n)_i}{(1)_i(b_1)_i\cdots(b_n)_i}t^i,
\]
where $(a)_i$ stands for the factorial function
\[
	(a)_i = a(a+1)\cdots(a+i-1) = \frac{\Gamma(a+i)}{\Gamma(a)}.
\]
Denoting by $\delta=td/dt$, we see that $x={}_{n+1}F_n$ satisf\/ies an $(n+1)$-th order linear dif\/ferential equation
\begin{gather}\label{Eq:Gen_HGE}
	[\delta(\delta+b_1-1)\cdots(\delta+b_n-1)-t(\delta+a_0)\cdots(\delta+a_n)]x = 0,
\end{gather}
which is called {\it the generalized hypergeometric equation} \cite{OTY}.
The equation \eqref{Eq:Gen_HGE} is of Fuchsian type with regular singular points at $t=0,1,\infty$ and the Riemann scheme
\[
	\left[\begin{array}{ccc}
		t=0& t=1& t=\infty\\
		0& 0& a_0\\
		1-b_1& 1& a_1\\
		\vdots& \vdots & \vdots\\
		1-b_{n-1}& n-1& a_{n-1}\\
		1-b_n& -\sum\limits_{i=1}^{n}(1-b_i)-\sum\limits_{i=0}^{n}a_i& a_n
	\end{array}\right].
\]
Note that ${}_{n+1}F_n$ includes the Gauss hypergeometric function as the case $n=1$.

In this article, we clarify a relationship between the system \eqref{Eq:CP6} and the function ${}_{n+1}F_n$.
For $n=1$ among them, the relationship between $P_{\rm{VI}}$ and the Gauss hypergeometric function is well known.
Under the system \eqref{Eq:CP6} of the case $n=1$, we consider a specialization $p=\eta=0$.
Then we obtain a Riccati equation
\[
	t(t-1)\frac{dq}{dt} = \alpha_1q^2 + \left\{(\alpha_3+\alpha_0)t-(\alpha_0+\alpha_1)\right\}q - \alpha_3t.
\]
Via a transformation of a dependent variable
\[
	q = \frac{t(1-t)}{\alpha_1}\frac{d}{dt}\log\{(t-1)^{\alpha_3}x(t)\},
\]
we obtain the Gauss hypergeometric equation
\[
	[\delta(\delta+\alpha_2+\alpha_3-1)-t(\delta+\alpha_1+\alpha_2+\alpha_3)(\delta+\alpha_3)]x = 0.
\]
The result of this article gives a natural extension of the above fact.
For general $n$, we consider a~specialization $p_1=\cdots=p_n=\eta=0$.
Then we obtain the generalized hypergeometric equation by a certain transformation of dependent variables.

We also discuss a degeneration structure of the system \eqref{Eq:CP6} derived from the conf\/luence of~${}_{n+1}F_n$.
The conf\/luent hypergeometric functions ${}_{n-r+1}F_n$ $(r=1,\ldots,n+1)$ are def\/ined by the power series
\[
	{}_{n-r+1}F_n\left[\begin{array}{c}a_r,\ldots,a_n\\b_1,\ldots,b_n\end{array};t\right] = \sum_{i=0}^{\infty}\frac{(a_r)_i\cdots(a_n)_i}{(b_1)_i\cdots(b_n)_i}t^i.
\]
The process of conf\/luence ${}_{n-r+2}F_n\to{}_{n-r+1}F_n$ is given by a replacement
\[
	t \to \varepsilon t,\qquad
	a_{r-1} \to \varepsilon^{-1},
\]
and taking a limit $\varepsilon\to0$.
We see that $x={}_{n-r+1}F_n$ satisfy the conf\/luent hypergeometric dif\/ferential equations
\begin{gather}\label{Eq:Conf_HGE}
	[\delta(\delta+b_1-1)\cdots(\delta+b_n-1)-t(\delta+a_r)\cdots(\delta+a_n)]x = 0.
\end{gather}
In this article, we propose a class of higher order Painlev\'{e} systems which admit particular solutions expressed in terms of ${}_{n-r+1}F_n$.

\begin{remark}
In this article, we study a higher order generalization of $P_{\rm{VI}}$.
On the other hand, for a multi-time generalization, it is known that the Garnier system admits a particular solution in terms of the Appell--Lauricella hypergeometric function $F_D$ \cite{IKSY}.
\end{remark}

This article is organized as follows.
In Section~\ref{Sec:Linear}, we derive a system of linear dif\/ferential equations from the system~\eqref{Eq:CP6} by a specialization $p_1=\cdots=p_n=\eta=0$.
In Section~\ref{Sec:HGE}, we give its fundamental solutions expressed in terms of the hypergeometric function ${}_{n+1}F_n$ in a~neighborhood of the singular point $t=0$.
In Section~\ref{Sec:Deg}, we discuss a degeneration structure of the system~\eqref{Eq:CP6} derived from the conf\/luence of ${}_{n+1}F_n$.

\section{Linear dif\/ferential equations}\label{Sec:Linear}

In this section, we derive a system of linear dif\/ferential equations from the system~\eqref{Eq:CP6} by a~specialization $p_1=\cdots=p_n=\eta=0$.

We f\/irst consider a {\it symmetric form} of \eqref{Eq:CP6} in order to derive a system of linear dif\/ferential equations.
Let $x_i$, $y_i$ $(i=0,\ldots,n)$ be dependent variables such that
\[
	t(1-t)\frac{d}{dt}\log x_n = \sum_{i=1}^{n}\left\{(q_i-1)(q_i-t)p_i+\alpha_{2i-1}q_i\right\} + t\alpha_{2n+1} - (t+1)\eta,
\]
and
\[
	x_{i-1} = \frac{x_nq_i}{t},\qquad
	y_{i-1} = \frac{tp_i}{x_n},\qquad  i=1,\ldots,n,\qquad
	y_n = -\frac{1}{x_n}\left(\sum_{j=1}^{n}q_jp_j+\eta\right).
\]
Then we obtain a Hamiltonian system of $(2n+2)$-th order
\begin{gather}\label{Eq:FS}
	\frac{dx_i}{dt} = \frac{\partial H}{\partial y_i},\qquad
	\frac{dy_i}{dt} = -\frac{\partial H}{\partial x_i},\qquad i=0,\ldots,n,
\end{gather}
with a Hamiltonian
\begin{gather*}
	H = \frac{1}{t}\sum_{i=0}^{n}\left\{\frac{1}{2}x_i^2y_i^2-\alpha_{2i+2}^{2n-2i-1}x_iy_i+\sum_{j=0}^{i-1}x_i(x_iy_i+\alpha_{2i+1})y_j\right\}\\
\phantom{H=}{} + \frac{1}{1-t}\sum_{i=0}^{n}\sum_{j=0}^{n}x_i(x_iy_i+\alpha_{2i+1})y_j,
\end{gather*}
where
\[
	\alpha_k^l = \left\{\begin{array}{ll}
		0, & l\in\mathbb{Z}_{<0},\\[4pt]
		\sum\limits_{i=k}^{k+l}\alpha_i, & l\in\mathbb{Z}_{\geq0}.
	\end{array}\right.
\]
The dependent variables $x_i$, $y_i$ and the f\/ixed parameter $\eta$ satisfy a relation
\[
	\sum_{i=0}^{n}x_iy_i + \eta = 0.
\]

\begin{remark}
The symmetric form \eqref{Eq:FS} is suggested by the Hamiltonian system given in Theorem~3.2 of~\cite{Su}.
Their relationship is given by
\[
	t = \frac{1}{t_1^{n+1}},\qquad
	x_i = \frac{w_{2i+1}}{t_1^{i-n+\rho_1+\kappa_{2n+1}-\kappa_0}},\qquad
	y_i = \frac{t_1^{i-n+\rho_1+\kappa_{2n+1}-\kappa_0}\varphi_{2i+1}}{n+1},
\]
and
\begin{gather*}
	\eta = \sum_{j=0}^{n}\frac{\rho_1+\kappa_{2i}-\kappa_{2i+1}}{n+1},\qquad
	\alpha_{2i} = \frac{1+\kappa_{2i-1}-2\kappa_{2i}+\kappa_{2i+1}}{n+1},\\
	\alpha_{2i+1} = \frac{\kappa_{2i}-2\kappa_{2i+1}+\kappa_{2i+2}}{n+1},
\end{gather*}
for $i=0,\ldots,n$.
\end{remark}

\begin{remark}
The system \eqref{Eq:FS}, or equivalently the system \eqref{Eq:CP6}, admits the af\/f\/ine Weyl group symmetry of type $A^{(1)}_{2n+1}$; see Appendix~\ref{Sec:Aff_Wyl_sym}.
\end{remark}

We can derive easily a system of linear dif\/ferential equations from the symmetric form by the specialization $y_0=\cdots=y_n=\eta=0$, which is equivalent to $p_1=\cdots=p_n=\eta=0$.
Let $E_{i,j}$ be the matrix unit def\/ined by
\[
	E_{i,j} = \left(\delta_{i,k}\delta_{j,l}\right)_{k,l=0}^{n}.
\]
For example, in the case $n=2$, it is explicitly given as
\[
	E_{1,1} = \begin{bmatrix}1&0\\0&0\end{bmatrix},\qquad
	E_{1,2} = \begin{bmatrix}0&1\\0&0\end{bmatrix},\qquad
	E_{2,1} = \begin{bmatrix}0&0\\1&0\end{bmatrix},\qquad
	E_{2,2} = \begin{bmatrix}0&0\\0&1\end{bmatrix}.
\]
Then we obtain

\begin{proposition}
The system \eqref{Eq:FS} admits a specialization
\[
	y_i = 0,\qquad i=0,\ldots,n,\qquad
	\eta = 0.
\]
Then a vector of the variables $\mathbf{x}={}^t(x_0,\ldots,x_n)$ satisfies a system of linear differential equations on $\mathbb{P}^1(\mathbb{C})$
\begin{gather}\label{Eq:HGE}
	\frac{d\mathbf{x}}{dt} = \left(\frac{A_0}{t}+\frac{A_1}{1-t}\right)\mathbf{x},
\end{gather}
with
\[
	A_0 = \sum_{i=0}^{n-1}\big(-\alpha_{2i+2}^{2n-2i-1}\big)E_{i,i} + \sum_{i=0}^{n-1}\sum_{j=i+1}^{n}\alpha_{2j+1}E_{i,j},\qquad
	A_1 = \sum_{i=0}^{n}\sum_{j=0}^{n}\alpha_{2j+1}E_{i,j}.
\]
Furthermore, the system \eqref{Eq:HGE} is of Fuchsian type with regular singular points at $t=0,1,\infty$.
The data of eigenvalues of its residue matrices is given as
\[\begin{array}{@{}lll}
	-\alpha_2^{2n-1}, \ldots, -\alpha_{2n}^1, 0& \text{at}& t=0,\vspace{1mm}\\
	0, \ldots, 0, \displaystyle-\sum_{i=0}^{n}\alpha_{2i+1}& \text{at}& t=1,\vspace{1mm}\\
	\alpha_1^{2n}, \ldots, \alpha_{2n-1}^2, \alpha_{2n+1}& \text{at}& t=\infty.
\end{array}\]
\end{proposition}

\begin{remark}
The system \eqref{Eq:FS} also admits a specialization
\[
	x_i = 0,\qquad i=0,\ldots,n-1,\qquad
	x_ny_n + \eta = 0,\qquad
	\eta - \alpha_{2n+1} = 0,
\]
which is equivalent to $q_1=\cdots=q_n=\eta-\alpha_{2n+1}=0$.
Then a vector of the variables $\mathbf{y}={}^t(y_0,\ldots,y_n)$ satisf\/ies a system of linear dif\/ferential equations
\[
	\frac{d\mathbf{y}}{dt} = \left(\frac{A_0}{t}+\frac{A_1}{1-t}\right)\mathbf{y},
\]
with
\begin{gather*}
	A_0  = \sum_{i=0}^{n-1}\alpha_{2i+2}^{2n-2i-1}E_{i,i} + \sum_{i=1}^{n-1}\sum_{j=0}^{i-1}(-\alpha_{2i+1})E_{i,j} + \sum_{j=0}^{n}\alpha_{2n+1}E_{n,j},\\
	A_1  = \sum_{i=0}^{n-1}\sum_{j=0}^{n}(-\alpha_{2i+1})E_{i,j} + \sum_{j=0}^{n}\alpha_{2n+1}E_{n,j}.
\end{gather*}
\end{remark}

We always assume that
\[
	\alpha_{2i}^{2j-1}\notin\mathbb{Z},\qquad
	\sum_{i=0}^{n}\alpha_{2i+1}\notin\mathbb{Z},\qquad
	\alpha_{2i-1}^{2j-1}\notin\mathbb{Z},\qquad i=1,\ldots,n,\quad j=1,\ldots,n-i+1.
\]
In the next section, we describe fundamental solutions of the system \eqref{Eq:HGE} in a neighborhood of the singular point $t=0$ explicitly.

\section[Hypergeometric function ${}_{n+1}F_n$]{Hypergeometric function $\boldsymbol{{}_{n+1}F_n}$}\label{Sec:HGE}

In this section, we give fundamental solutions of the system \eqref{Eq:HGE} expressed in terms of the hypergeometric function ${}_{n+1}F_n$ in a neighborhood of the singular point $t=0$.

For each $k=0,\ldots,n$, we consider a gauge transformation
\[
	\mathbf{x}^k = t^{\alpha_{2k+2}^{2n-2k-1}}\left(\sum_{i=0}^{n-k-1}t^{-1}E_{i,i+k+1}+\sum_{i=n-k}^{n}E_{i,i-n+k}\right)\mathbf{x}.
\]
Then the system \eqref{Eq:HGE} is transformed into
\begin{gather}\label{Eq:HGE_series}
	\frac{d\mathbf{x}^k}{dt} = \left(\frac{A_0^k}{t}+\frac{A_1^k}{1-t}\right)\mathbf{x}^k,
\end{gather}
with
\begin{gather*}
	A_0^k  = \sum_{i=0}^{n-1}(-\alpha_{2k+2i+4}^{2n-2i-1})E_{i,i} + \sum_{i=0}^{n-1}\sum_{j=i+1}^{n}\alpha_{2j+2k+3}E_{i,j},\\
	A_1^k  = \sum_{i=0}^{n}\sum_{j=0}^{n}\alpha_{2j+2k+3}E_{i,j}.
\end{gather*}
Recall that indices of the f\/ixed parameters $\alpha_i$ are congruent modulo $2n+2$, from which we have $\alpha_{2k+2n+2}^1=\alpha_{2k}^1$.
We also consider a formal power series of $\mathbf{x}^k$ at $t=0$
\[
	\mathbf{x}^k = \sum_{i=0}^{\infty}\mathbf{x}_i^kt^i.
\]
Then the system \eqref{Eq:HGE_series} implies
\begin{gather}
	 A_0^k\mathbf{x}_0^k = \mathbf{0},\nonumber\\
	 \big\{A_0^k-(i+1)I\big\}\mathbf{x}_{i+1}^k = \big(A_0^k-A_1^k-iI\big)\mathbf{x}_i^k,\qquad i\in\mathbb{Z}_{\geq0},\label{Eq:HGE_series_rec}
\end{gather}
where $I$ stands for the identity matrix.
The matrices $A_0^k$ and $A_0^k-(i+1)I$ are of rank $n$ and $n+1$, respectively.
It follows that the recurrence formula~\eqref{Eq:HGE_series_rec} admits one parameter family of solutions.

For each $k=0,\ldots,n$, we can show that a sequence of vectors
\[
	\mathbf{x}_i^k = \begin{bmatrix}\prod\limits_{j=0}^{n-1}\frac{(\alpha_{2k-2j+1}^{2j})_{i+1}}{(\alpha_{2k-2j}^{2j+1})_{i+1}}\cdot
\frac{(\alpha_{2k+3}^{2n})_i}{(\alpha_{2k+2}^{2n+1})_i}\vspace{1mm}\\
\prod\limits_{j=0}^{n-2}\frac{(\alpha_{2k-2j+1}^{2j})_{i+1}}{(\alpha_{2k-2j}^{2j+1})_{i+1}}\cdot\frac{(\alpha_{2k+5}^{2n-2})_i
(\alpha_{2k+3}^{2n})_i}{(\alpha_{2k+4}^{2n-1})_i(\alpha_{2k+2}^{2n+1})_i}\\
\vdots\\
\frac{(\alpha_{2k+1})_{i+1}}{(\alpha_{2k}^1)_{i+1}}\prod\limits_{j=0}^{n-1}\frac{(\alpha_{2k+2j+3}^{2n-2j})_i}{(\alpha_{2k+2j+2}^{2n-2j+1})_i}
\vspace{1mm}\\
\prod\limits_{j=0}^{n}\frac{(\alpha_{2k+2j+3}^{2n-2j})_i}{(\alpha_{2k+2j+2}^{2n-2j+1})_i}\end{bmatrix},\qquad i\in\mathbb{Z}_{\geq0},
\]
satisf\/ies the recurrence formula \eqref{Eq:HGE_series_rec} by a direct computation.
Note that $\alpha_{2k+2}^{2n+1}=1$.
Therefore we arrive at

\begin{theorem}
On a domain $|t|<1$, the system \eqref{Eq:HGE} admits fundamental solutions
\[
	\mathbf{x} = t^{-\alpha_{2k+2}^{2n-2k-1}}\begin{bmatrix}f^{k,k}\\\vdots\\f^{k,0}\\tf^{k,n}\\\vdots\\tf^{k,k+1}\end{bmatrix},\qquad k=0,\ldots,n,
\]
where
\[
	f^{k,l} = \prod_{i=1}^{l}\frac{\alpha_{2k-2i+3}^{2i-2}}{\alpha_{2k-2i+2}^{2i-1}}\cdot{}_{n+1}F_n
\left[\begin{array}{c}a_0,\ldots,a_n\\b_1,\ldots,b_n\end{array};t\right],
\]
with
\begin{alignat*}{4}
	& a_0 = \alpha_{2k-2n+1}^{2n}, && && & \\
	& a_i = 1+\alpha_{2k-2i+3}^{2i-2}, \qquad && b_i = 1+\alpha_{2k-2i+2}^{2i-1}, \qquad && i=1,\ldots,l,& \\
	& a_i = \alpha_{2k-2i+3}^{2i-2},\qquad && b_i = \alpha_{2k-2i+2}^{2i-1},\qquad && i=l+1,\ldots,n. &
\end{alignat*}
\end{theorem}

\begin{corollary}
If the vector $\mathbf{x}={}^t(x_0,\ldots,x_n)$ satisfies the system \eqref{Eq:HGE}, each component $x_i$ satisfies the generalized hypergeometric equation \eqref{Eq:Gen_HGE} with
\begin{alignat*}{4}
	& a_0 = \alpha_{1}^{2n},&& && & \\
	& a_j = 1+\alpha_{2n-2j+3}^{2j-2},\qquad && b_j = 1+\alpha_{2n-2j+2}^{2j-1},\qquad  && j=1,\ldots,n-i,&\\
	& a_j = \alpha_{2n-2j+3}^{2j-2},\qquad && b_j = \alpha_{2n-2j+2}^{2j-1}, \qquad && j=n-i+1,\ldots,n.&
\end{alignat*}
\end{corollary}

\begin{remark}
The system \eqref{Eq:HGE} have been already studied by Okubo--Takano--Yoshida~\cite{OTY}.
They considered the Fuchsian dif\/ferential equation of Okubo type and obtained its fundamental solutions at singular points $t=0,1$.
\end{remark}

\section{Degeneration structure}\label{Sec:Deg}

In this section, we discuss a degeneration structure of the system \eqref{Eq:CP6} derived from the conf\/luence of ${}_{n+1}F_n$.

For each $r=1,\ldots,n+1$, we consider a Hamiltonian system
\[
	{}_{n-r+1}\mathcal{H}_n:\ \
	\frac{dx_i}{dt} = \frac{\partial H}{\partial y_i},\qquad
	\frac{dy_i}{dt} = -\frac{\partial H}{\partial x_i},\qquad i=0,\ldots,n,
\]
with a Hamiltonian
\begin{gather*}
	tH = \sum_{i=0}^{n}\frac{1}{2}x_iy_i\big(x_iy_i-2\alpha_{2i+2}^{2n-2i-1}\big)
	+ \sum_{i=0}^{r-2}x_{i+1}y_i + \sum_{i=r-1}^{n}\!\left\{tx_0+\sum_{j=i+1}^{n}x_j(x_jy_j+\alpha_{2j+1})\right\}y_i.
\end{gather*}
Here $\alpha_i$ $(i=0,\ldots,2n+1)$ and $\eta$ are f\/ixed parameters satisfying
\[
	\alpha_{2i} = 0,\quad i=0,\ldots,r-1,\qquad
	\sum_{j=0}^{n}\alpha_{2j+1} + \sum_{j=r}^{n}\alpha_{2j} = 1,
\]
and
\[
	\sum_{j=0}^{n}x_jy_j + \eta = 0.
\]
Note that
\[
	\alpha_{2i+2}^{2n-2i-1} = \sum_{j=2i+2}^{2n+1}\alpha_j = \sum_{j=i+1}^{n}\alpha_{2j+1} + \sum_{j=\max (r,i+1)}^{n}\alpha_{2j}.
\]
The system ${}_{n-r+1}\mathcal{H}_n$ is obtained from ${}_{n-r+2}\mathcal{H}_n$ by a replacement
\begin{gather*}
	t \to \varepsilon t,\qquad
	\alpha_{2r-2} \to -\varepsilon^{-1},\qquad
	\alpha_{2r-1} \to \alpha_{2r-1} + \varepsilon^{-1},\\
	x_i \to \varepsilon^{-1}x_i,\qquad
	y_i \to \varepsilon y_i, \qquad i=0,\ldots,r-2.
\end{gather*}
and taking a limit $\varepsilon\to0$, where ${}_{n+1}\mathcal{H}_n$ stands for the system \eqref{Eq:FS}.

\begin{remark}
Such degenerate systems also can be rewritten into the Hamiltonian systems in terms of the canonical coordinates.
We give their explicit formulas for $n=1$ and $n=2$ in Appendix~\ref{Sec:Cano_Ham}.
\end{remark}

The system ${}_{n-r+1}\mathcal{H}_n$ admits a specialization
\[
	y_i = 0,\quad i=0,\ldots,n,\qquad
	\eta = 0.
\]
Then a vector of the variables $\mathbf{x}={}^t(x_0,\ldots,x_n)$ satisf\/ies a system of linear dif\/ferential equations
\[
	{}_{n-r+1}\mathcal{L}_n:\ \
	\frac{d\mathbf{x}}{dt} = \left(\frac{A_0}{t}+A_1\right)\mathbf{x},
\]
with
\begin{gather*}
	A_0  = \sum_{i=0}^{n-1}\big(-\alpha_{2i+2}^{2n-2i-1}\big)E_{i,i} + \sum_{i=0}^{r-2}E_{i,i+1} + \sum_{i=r-1}^{n-1}\sum_{j=i+1}^{n}\alpha_{2j+1}E_{i,j},\\
	A_1  = \sum_{i=r-1}^{n}E_{i,0}.
\end{gather*}
Note that ${}_{n-r+1}\mathcal{L}_n$ is obtained from ${}_{n-r+2}\mathcal{L}_n$ through the above process of conf\/luence.

In a similar manner as Section \ref{Sec:HGE}, we arrive at

\begin{theorem}
On a domain $|t|<1$, the system ${}_{n-r+1}\mathcal{L}_n$ admits fundamental solutions
\[
	\mathbf{x} = t^{-\alpha_{2k+2}^{2n-2k-1}}\begin{bmatrix}f^{k,k}_r\\\vdots\\f^{k,0}_r\\tf^{k,n}_r\\\vdots\\tf^{k,k+1}_r\end{bmatrix},\qquad k=0,\ldots,n,
\]
where
\[
	f^{k,l}_r = \prod_{\substack{1\leq i\leq l\\\mathrm{mod}[k-i+1,n+1]\geq r}}\alpha_{2k-2i+3}^{2i-2}\prod_{1\leq i\leq l}\frac{1}{\alpha_{2k-2i+2}^{2i-1}}\cdot{}_{n-r+1}F_n\left[\begin{array}{c}a_r,\ldots,a_n\\b_1,\ldots,b_n\end{array};t\right],
\]
and
\[
	\mathrm{mod}[i,n+1] = i - m(n+1)\qquad \text{for}\quad m(n+1)\leq i<(m+1)(n+1).
\]
Here the parameters $a_r,\ldots,a_n$ are given by
\[
	a_i = \alpha_{2r-2i-1}^{2k-2r+2i+2}, \qquad i=r,\ldots,n,
\]
for $k+1\leq r$ and $l<k+2$;
\begin{alignat*}{3}
	& a_i = 1+\alpha_{2r-2i-1}^{2k-2r+2i+2}, \qquad && i=r,\ldots,r-k+l-2,& \\
	& a_i = \alpha_{2r-2i-1}^{2k-2r+2i+2}, \qquad && i=r-k+l-1,\ldots,n,&
\end{alignat*}
for $k+1\leq r$ and $k+2\leq l$;
\begin{alignat*}{3}
	& a_i = \alpha_{2r-2i-1}^{2k-2r+2i+2}, \qquad && i=r,\ldots,n+r-k-1, &\\
	& a_i = 1+\alpha_{2r-2i-1}^{2k-2r+2i+2}, \qquad && i=n+r-k,\ldots,n+r-k+l-1,& \\
	& a_i = \alpha_{2r-2i-1}^{2k-2r+2i+2}, \qquad && i=n+r-k+l,\ldots,n, &
\end{alignat*}
for $r<k+1$ and $l<k-r+1$;
\begin{alignat*}{3}
	& a_i = \alpha_{2r-2i-1}^{2k-2r+2i+2}, \qquad && i=r,\ldots,n+r-k-1,& \\
	& a_i = 1+\alpha_{2r-2i-1}^{2k-2r+2i+2},\qquad && i=n+r-k,\ldots,n,&
\end{alignat*}
for $r<k+1$ and $k-r+1\leq l<k+2$;
\begin{alignat*}{3}
	& a_i = 1+\alpha_{2r-2i-1}^{2k-2r+2i+2}, \qquad && i=r,\ldots,r-k+l-2,&\\
	& a_i = \alpha_{2r-2i-1}^{2k-2r+2i+2}, \qquad && i=r-k+l-1,\ldots,n+r-k-1,& \\
	& a_i = 1+\alpha_{2r-2i-1}^{2k-2r+2i+2},\qquad && i=n+r-k,\ldots,n,&
\end{alignat*}
for $r<k+1$ and $k+2\leq l$.
The parameters $b_1,\ldots,b_n$ are given by
\begin{alignat*}{3}
	& b_i = 1+\alpha_{2k-2i+2}^{2i-1},\qquad && i=1,\ldots,l,&\\
	& b_i = \alpha_{2k-2i+2}^{2i-1},\qquad && i=l+1,\ldots,n.&
\end{alignat*}
\end{theorem}

\begin{corollary}
If the vector $\mathbf{x}={}^t(x_0,\ldots,x_n)$ satisfies the system ${}_{n-r+1}\mathcal{L}_n$, each component $x_i$ satisfies the confluent hypergeometric equation \eqref{Eq:Conf_HGE} with
\[
	a_j = 1+\alpha_{2r-2j-1}^{2n-2r+2j+2}, \qquad j=r,\ldots,n,
\]
for $i\leq r-1$,
\begin{alignat*}{3}
	& a_j = 1+\alpha_{2r-2j-1}^{2n-2r+2j+2}, \qquad && j=r,\ldots,n+r-i-1,& \\
	& a_j = \alpha_{2r-2j-1}^{2n-2r+2j+2},\qquad && j=n+r-i,\ldots,n,&
\end{alignat*}
for $r-1<i$ and
\begin{alignat*}{3}
	& b_j = 1+\alpha_{2n-2j+2}^{2j-1}, \qquad && j=1,\ldots,n-i,&\\
	& b_j = \alpha_{2n-2j+2}^{2j-1}, \qquad && j=n-i+1,\ldots,n,
\end{alignat*}
for any $i$.
\end{corollary}

\appendix

\section{Canonical Hamiltonian system}\label{Sec:Cano_Ham}

The systems ${}_{n-r+1}\mathcal{H}_n$ can be rewritten into the Hamiltonian systems in terms of canonical coordinates.
In this section, we give their explicit formulas for $n=1$ and $n=2$.
Note that ${}_{3-r}\mathcal{H}_2$ appear in the classif\/ication of the fourth order Painlev\'{e} type dif\/ferential equations \cite{Sa}.

\subsection[Case $n=1$, $r=1$]{Case $\boldsymbol{n=1}$, $\boldsymbol{r=1}$}

Under the system ${}_1\mathcal{H}_1$, we take canonical coordinates
\[
	q = \frac{x_0}{x_1},\qquad
	p = -\frac{x_1(x_1y_1+\alpha_3)}{x_0}.
\]
Via a transformation of the independent variable $t\to-t$, we obtain a Hamiltonian system
\[
	\frac{dq}{dt} = \frac{\partial H}{\partial p},\qquad
	\frac{dp}{dt} = -\frac{\partial H}{\partial q},
\]
with a Hamiltonian
\[
	tH = q(q-1)p(p+t) - qp(\eta+\alpha_2-\alpha_3) + (\eta-\alpha_3)p + t\alpha_3q.
\]
It is equivalent to the f\/ifth Painlev\'{e} equation.

\subsection[Case $n=1$, $r=2$]{Case $\boldsymbol{n=1}$, $\boldsymbol{r=2}$}

Under the system ${}_0\mathcal{H}_1$, we take canonical coordinates
\[
	q = \frac{x_1}{x_0},\qquad
	p = x_0y_1.
\]
Then we obtain a Hamiltonian system
\[
	\frac{dq}{dt} = \frac{\partial H}{\partial p},\qquad
	\frac{dp}{dt} = -\frac{\partial H}{\partial q},
\]
with a Hamiltonian
\[
	tH = q^2p(p-1) + (\eta+\alpha_3)qp + tp - \eta q.
\]
It is equivalent to the third Painlev\'{e} equation.

\subsection[Case $n=2$, $r=1$]{Case $\boldsymbol{n=2}$, $\boldsymbol{r=1}$}

Under the system ${}_2\mathcal{H}_2$, we take canonical coordinates
\[
	q_1 = \frac{x_0}{x_1},\qquad
	p_1 = -\frac{x_1(x_1y_1+\alpha_3)}{x_0},\qquad
	q_2 = \frac{x_0}{x_2},\qquad
	p_2 = -\frac{x_2(x_2y_2+\alpha_5)}{x_0}.
\]
Via a transformation of the independent variable $t\to-t$, we obtain a Hamiltonian system
\[
	\frac{dq_i}{dt} = \frac{\partial H}{\partial p_i},\qquad
	\frac{dp_i}{dt} = -\frac{\partial H}{\partial q_i},\qquad i=1,2,
\]
with a Hamiltonian
\begin{gather*}
	tH  = q_1(q_1-1)p_1(p_1+t) - (\eta+\alpha_2-\alpha_3-\alpha_5)q_1p_1 + (\eta-\alpha_3-\alpha_5)p_1\\
\phantom{tH  =}{} + \alpha_3tq_1 + (q_1-1)p_1q_2p_2+ (q_1-1)(q_1p_1+\alpha_3)p_2\\
\phantom{tH  =}{} + q_2(q_2-1)p_2(p_2+t) - (\eta+\alpha_2+\alpha_4-\alpha_5)q_2p_2 + (\eta-\alpha_5)p_2 + \alpha_5tq_2.
\end{gather*}

\subsection[Case $n=2$, $r=2$]{Case $\boldsymbol{n=2}$, $\boldsymbol{r=2}$}

Under the system ${}_1\mathcal{H}_2$, we take canonical coordinates
\[
	q_1 = -\frac{x_1}{x_0},\qquad
	p_1 = 1 - x_0y_1,\qquad
	q_2 = -\frac{x_2}{x_0},\qquad
	p_2 = -x_0y_2.
\]
Via a transformation of the independent variable $t\to-t$, we obtain a Hamiltonian system
\[
	\frac{dq_i}{dt} = \frac{\partial H}{\partial p_i},\qquad
	\frac{dp_i}{dt} = -\frac{\partial H}{\partial q_i},\qquad i=1,2,
\]
with a Hamiltonian
 \begin{gather*}
	tH  = q_1^2p_1(p_1-1) + (\eta+\alpha_3)q_1p_1 + tp_1 - \alpha_3q_1 + q_1p_1q_2p_2 + p_1q_2(q_2p_2+\alpha_5)\\
\phantom{tH  =}{} + q_2^2p_2(p_2-1) + (\eta+\alpha_3+\alpha_4+\alpha_5)q_2p_2 + tp_2 - \alpha_5q_2.
\end{gather*}

\subsection[Case $n=2$, $r=3$]{Case $\boldsymbol{n=2}$, $\boldsymbol{r=3}$}

Under the system ${}_0\mathcal{H}_2$, we take canonical coordinates
\[
	q_1 = -\frac{x_1}{x_0},\qquad
	p_1 = 1 - x_0y_1,\qquad
	q_2 = -\frac{x_2}{x_0},\qquad
	p_2 = -x_0y_2.
\]
Via a transformation of the independent variable $t\to-t$, we obtain a Hamiltonian system
\[
	\frac{dq_i}{dt} = \frac{\partial H}{\partial p_i},\qquad
	\frac{dp_i}{dt} = -\frac{\partial H}{\partial q_i},\qquad i=1,2,
\]
with a Hamiltonian
 \begin{gather*}
	tH  = q_1^2p_1(p_1-1) + (\eta+\alpha_3)q_1p_1 - \alpha_3q_1 + q_1p_1q_2p_2 + p_1q_2\\
\phantom{tH  =}{}  + q_2^2p_2^2 + (\eta+\alpha_3+\alpha_5)q_2p_2 + tp_2 - q_2.
\end{gather*}

\section{Af\/f\/ine Weyl group symmetry}\label{Sec:Aff_Wyl_sym}

The system \eqref{Eq:FS} admits the af\/f\/ine Weyl group symmetry of type $A^{(1)}_{2n+1}$.
In this section, we describe its action on the dependent variables and parameters.

Recall that the af\/f\/ine Weyl group of type $A^{(1)}_{2n+1}$ is generated by the transformations $r_i$ $(i=0,\ldots,2n+1)$ with the fundamental relations
 \begin{alignat*}{3}
	& r_i^2=1,\qquad && i=0,\ldots,2n+1,& \\
	& (r_ir_j)^{2-a_{i,j}}=0,\qquad && i,j=0,\ldots,2n+1, \quad i\neq j,&
\end{alignat*}
where
 \begin{alignat*}{3}
	& a_{i,i}=2, \qquad &&  i=0,\ldots,2n+1,& \\
	& a_{i,i+1}=a_{2n+1,0}=a_{i+1,i}=a_{0,2n+1}=-1, \qquad && i=0,\ldots,2n,& \\
	& a_{i,j}=0, \qquad && \text{otherwise}. &
\end{alignat*}

We def\/ine the Poisson structure by
\[
	\{x_i,y_j\} = -\delta_{i,j},\qquad i,j=0,\ldots,n.
\]
Then the Hamiltonian system \eqref{Eq:FS} is invariant under the birational transformations $r_0,\ldots,r_{2n+1}$ def\/ined by
 \begin{gather*}
	r_0(x_j) = t^{-\alpha_0}x_j,\qquad
	r_0(y_j) = t^{\alpha_0}\left(y_j+\frac{\alpha_0}{x_n-tx_0}\{x_n-tx_0,y_j\}\right),\\
	r_{2i+1}(x_j) = x_j + \frac{\alpha_{2i+1}}{y_i}\{y_i,x_j\},\qquad
	r_{2i+1}(y_j) = y_j,\qquad i=0,\ldots,n-1,\\
	r_{2i}(x_j) = x_j,\qquad
	r_{2i}(y_j) = y_j + \frac{\alpha_{2i}}{x_{i-1}-x_i}\{x_{i-1}-x_i,y_j\},\qquad i=1,\ldots,n,\\
	r_{2n+1}(x_j) = t^{\alpha_{2n+1}}\left(x_j+\frac{\alpha_{2n+1}}{y_n}\{y_n,x_j\}\right),\qquad
	r_{2n+1}(y_j) = t^{-\alpha_{2n+1}}y_j,
\end{gather*}
for $j=0,\ldots,n$ and
\[
	r_i(\alpha_j) = \alpha_j - a_{i,j}\alpha_i,\qquad
	r_i(\eta) = \eta + (-1)^i\alpha_i,\qquad i,j=0,\ldots,2n+1.
\]
The group of symmetries $\langle r_0,\ldots,r_{2n+1}\rangle$ is isomorphic to the af\/f\/ine Weyl group of type $A^{(1)}_{2n+1}$.

\subsection*{Acknowledgements}

The author would like to express his gratitude to Mr. Masaomi Miyamoto of Kobe University for fruitful discussion.
The author is also grateful to Professors Masatoshi Noumi, Hidetaka Sakai, Teruhisa Tsuda and Yasuhiko Yamada for helpful comments and advices.

\pdfbookmark[1]{References}{ref}
\LastPageEnding

\end{document}